\documentclass[a4paper]{jpconf}
\usepackage{graphicx}
\usepackage{slashed}
\usepackage{amsmath}
\usepackage{color}

\newcommand{\unit}[1]{\ensuremath{\mathrm{~#1}}}
\newcommand{\wjets}[0]{\mathrm{W+jets}}
\newcommand{\particle}[1]{\ensuremath{#1}}
\newcommand{\ttbar}[0]{\ensuremath{\mathrm{t\bar{t}}}}
\newcommand{\costheta}[0]{\cos\theta_{\mathrm{l,q}}^{\mathrm{(top)}}}
\newcommand{\pT}[0]{\ensuremath{p_\mathrm{T}}}
\newcommand{\mtw}[0]{\ensuremath{M_\mathrm{T}(W)}}
\newcommand{\met}[0]{\ensuremath{\slashed E_\mathrm{T}}}
\newcommand{\BDT}[0]{\ensuremath{\mathrm{BDT}}}

\begin{document}

\title{Measurement of Top-Quark Polarization in t-channel Single-Top Production}

\author{Matthias Komm}

\address{Centre for Cosmology, Particle Physics and Phenomenology, Universit\'e catholique de Louvain, 1348 Louvain-la-Neuve, Belgium}

\ead{Matthias.Komm@CERN.ch}

\begin{abstract}
The measurement of the top quark polarization, sensitive to the electroweak coupling structure, in t-channel single-top production is presented. Events are analyzed corresponding to an integrated luminosity of approximately $20\unit{fb^{-1}}$ recorded with the CMS detector during pp collisions at $\sqrt{s}=8\unit{TeV}$. By requiring one isolated lepton (muon or electron), two jets, and missing transverse energy, an angular asymmetry, sensitive to the  polarization of the top quark, is reconstructed in the top-quark rest frame. The corresponding angular asymmetry at parton level is inferred from data in a phase space with enhanced single-top t-channel candidates through unfolding. Remaining background contributions are estimated through a ML-fit and subtracted. A polarization of $P_{t}=0.82\pm0.12\mathrm{~(stat.)}\pm0.32\mathrm{~(syst.)}$ is measured assuming a spin-analyzing power of the charged lepton stemming from the top decay of $100\%$.
\end{abstract}

\section{Introduction}
In the theory of particle physics, the Standard Model (SM), electroweak interactions between fermions via charged currents are maximally parity violating. Only left-handed fermions (or right-handed anti-fermions) can couple to W bosons through a V-A coupling structure.

The top quark offers an unique possibility amongst all quarks to probe this prediction because of its very short lifetime below the hadronization time scale. Therefore, its spin orientation stays encoded in the angular distribution of its decay products.

An observable sensitive to the electroweak top quark coupling structure is given in t-channel single top-quark production by the forward-backward asymmetry
\begin{equation}
\label{eq:asym}
A=\frac{N(\costheta>0)-N(\costheta<0)}{N(\costheta>0)+N(\costheta<0)}=\frac{1}{2}P_{t}\alpha_{l} 
\end{equation}
in the top-quark rest frame, where $\theta_{\mathrm{l,q}}^{\mathrm{(top)}}$ denotes the angle between the lepton and the light ($\particle{u}$, $\particle{d}$, $\particle{s}$, $\particle{c}$) quark which may also be referred to as spectator quark. The polarization, $P_{t}$, denotes the alignment of the top-quark spin with the light-quark momentum and the spin-analyzing power, $\alpha_{l}$, quantifies the alignment of lepton with the top-quark spin. Theoretical calculations show that the particular V-A structure leads to a high polarization, $P_{t}=0.98$, and spin analyzing power $\alpha_{l}=1$~\cite{bernreuther}.

Beyond verifying the theoretical prediction, potential new particles or interactions beyond the SM might preferably occur in the top quark sector because the top quark is the heaviest known elementary particle, its measured mass of $173.34\pm 0.76\unit{GeV}$~\cite{topmass} is close to the electroweak symmetry breaking scale, and it gives major contributions to the Higgs self-energy. New physics influencing the electroweak top quark production and decay vertex coupling structure can be characterized in their low energy limit through an effective field theory (EFT)~\cite{jaaswpol}. Beyond this EFT scenario, the polarization is furthermore sensitive to additional anomalous couplings which are only possible in the production such as contact-interactions~\cite{fabian}.

This note shorty summarizes the preliminary result of the top quark polarization measurement which can be found here~\cite{stpol}.

\section{Event selection}
The measurement is based on data recorded with the CMS detector~\cite{cms} in 2012 during pp collisions at a center-of-mass energy of $\sqrt{s}=8\unit{TeV}$ which corresponds to an integrated luminosity of $19.7\pm0.5\unit{fb^{-1}}$. An isolated and central muon (or electron) with $\pT>24~(27)\unit{GeV}$ is requested to have fired the trigger.

Single top t-channel events and background processes ($\wjets$, $\ttbar$-pair production, Drell-Yan, and Di-Boson production) have been simulated using various event generators (\textsc{PowHeg}, \textsc{Pythia}, \textsc{MadGraph}, \textsc{Sherpa}, \textsc{CompHep}) which were interfaced with \textsc{Tauola} for $\tau$ decays and \textsc{Pythia} for hadronization and matching. The simulated events are passed through the \textsc{Geant4}-based CMS detector simulation and its reconstruction software. The shape and rate of the QCD multijet processes background is estimated from data.

Events in the signal region are required to contain one isolated muon (or electron) with $\pT>26~(30)\unit{GeV}$ and $|\eta|<2.1~(2.5)$, two jets with $\pT>40\unit{GeV}$ where one is b-tagged using the track-counting algorithm~\cite{tc}, and missing transverse energy. To reject large contribution from QCD multijet processes an additional selection on the transverse W-boson mass of $\mtw>40\unit{GeV}$ for the muon channel and on the missing transverse energy $\met>50\unit{GeV}$ for the electron channel is applied.

Candidate events in a signal-enhanced phase space are singled-out by applying an additional selection on the output of a Boosted Decision Tree (BDT) which was trained to discriminate signal against $\wjets$ and $\ttbar$ events. The BDT input variables were carefully selected to be uncorrelated to $\costheta$.

The contributions of signal and background processes in the analysis phase space are estimated by a maximum likelihood (ML) fit in the BDT output distribution using the simulated events as templates.

\begin{figure}[h]
\begin{center}
\begin{minipage}{7cm}
\includegraphics[height=6.0cm]{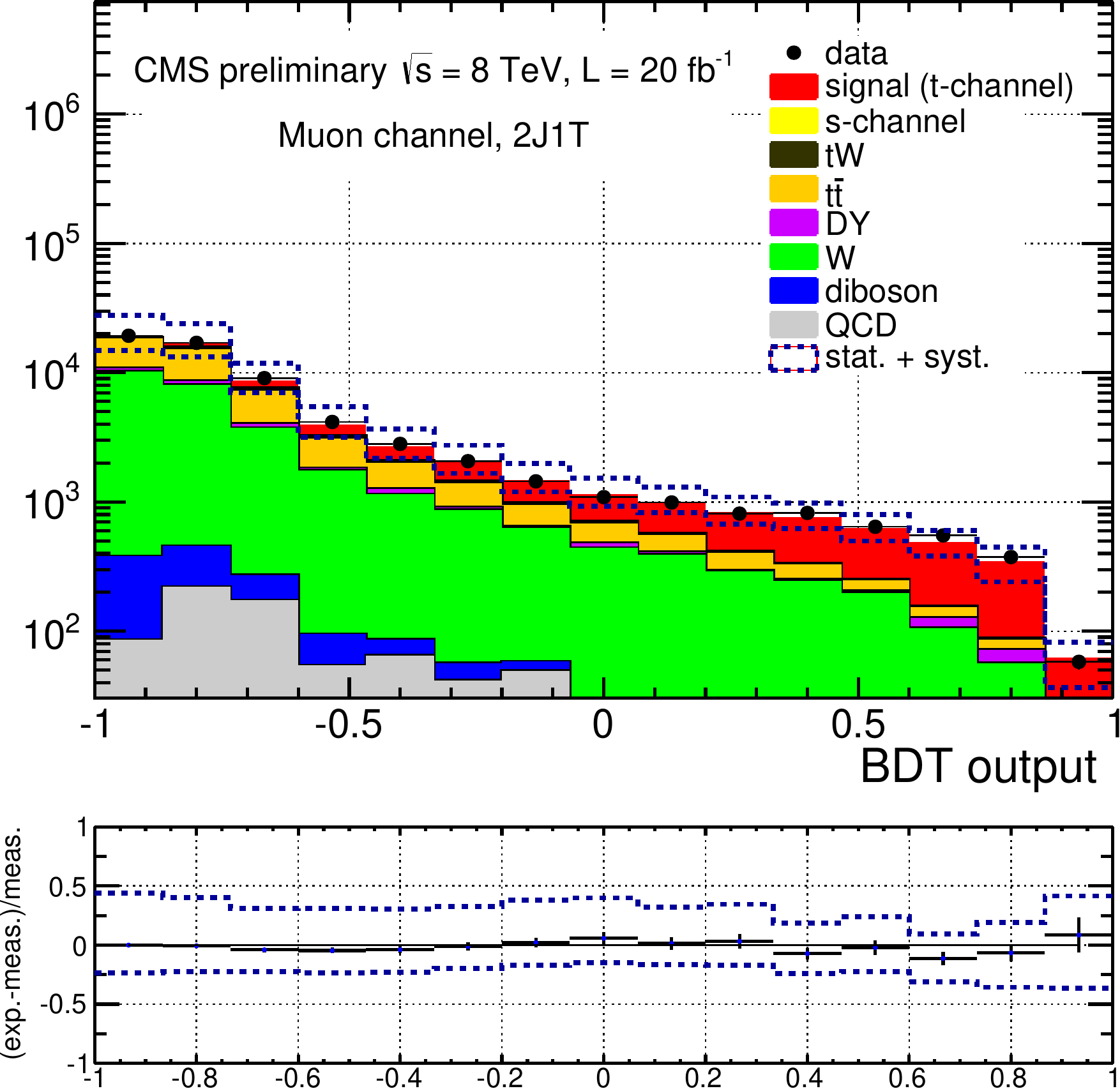}
\center (a) muon channel
\end{minipage}\hspace{1cm}%
\begin{minipage}{7cm}
\includegraphics[height=6.0cm]{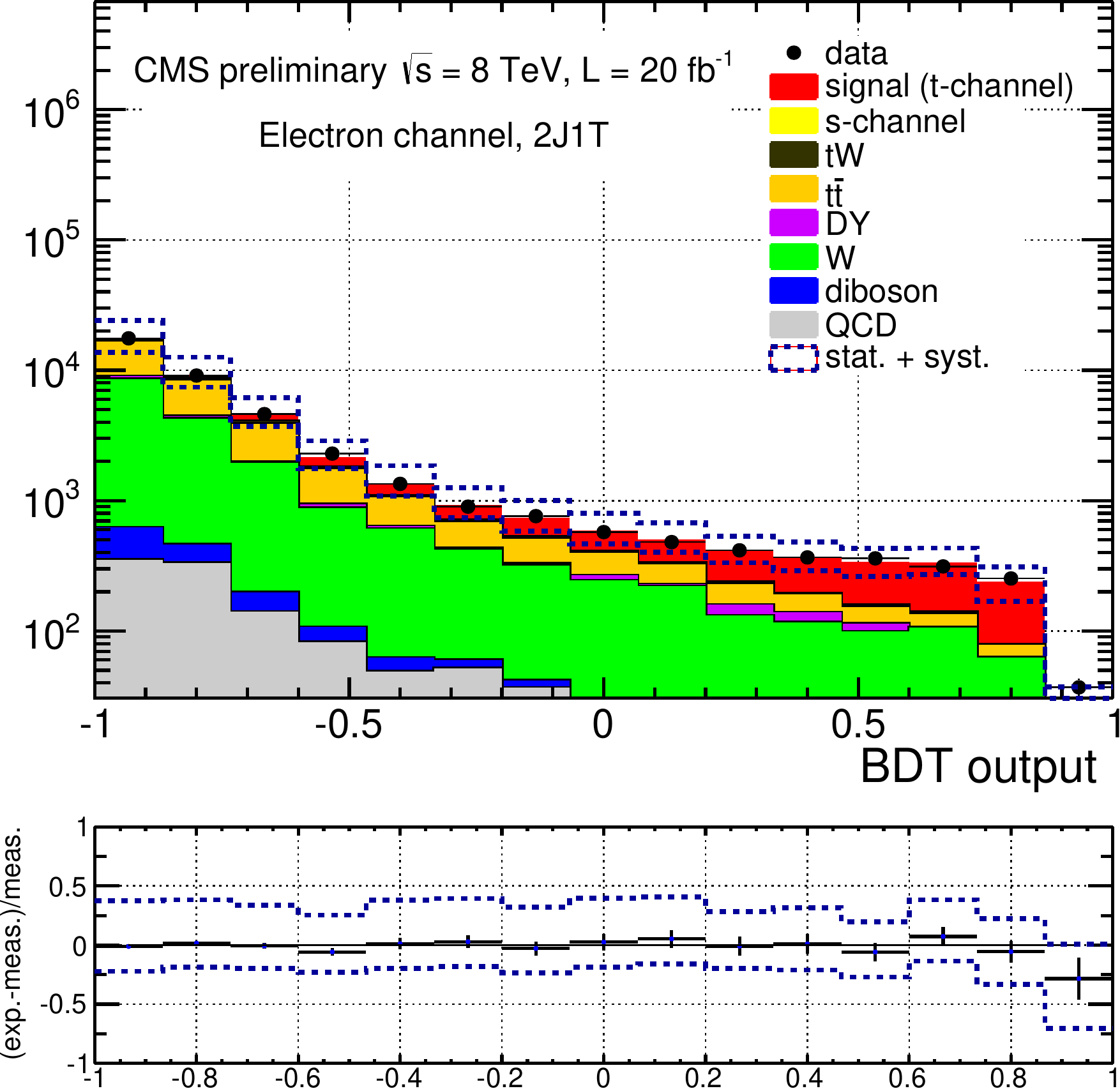}
\center (b) electron channel
\end{minipage} 
\caption{\label{fig:bdt}Output of the BDT trained to select a signal-enhanced phase space.}
\end{center}
\end{figure}

Figure~\ref{fig:bdt} show the BDT output after scaling the signal and background contributions to their ML-fit estimate. To verify the modeling of the simulated backgrounds, the BDT distribution is checked in the $\wjets$-dominated control region (2 jets, 0 b-tagged) and $\ttbar$-dominated control regions (3 jets, 1 or 2 b-tagged). No mis-modeling was found when comparing to data.

\begin{figure}[h]
\begin{center}
\begin{minipage}{7cm}
\includegraphics[height=6.0cm]{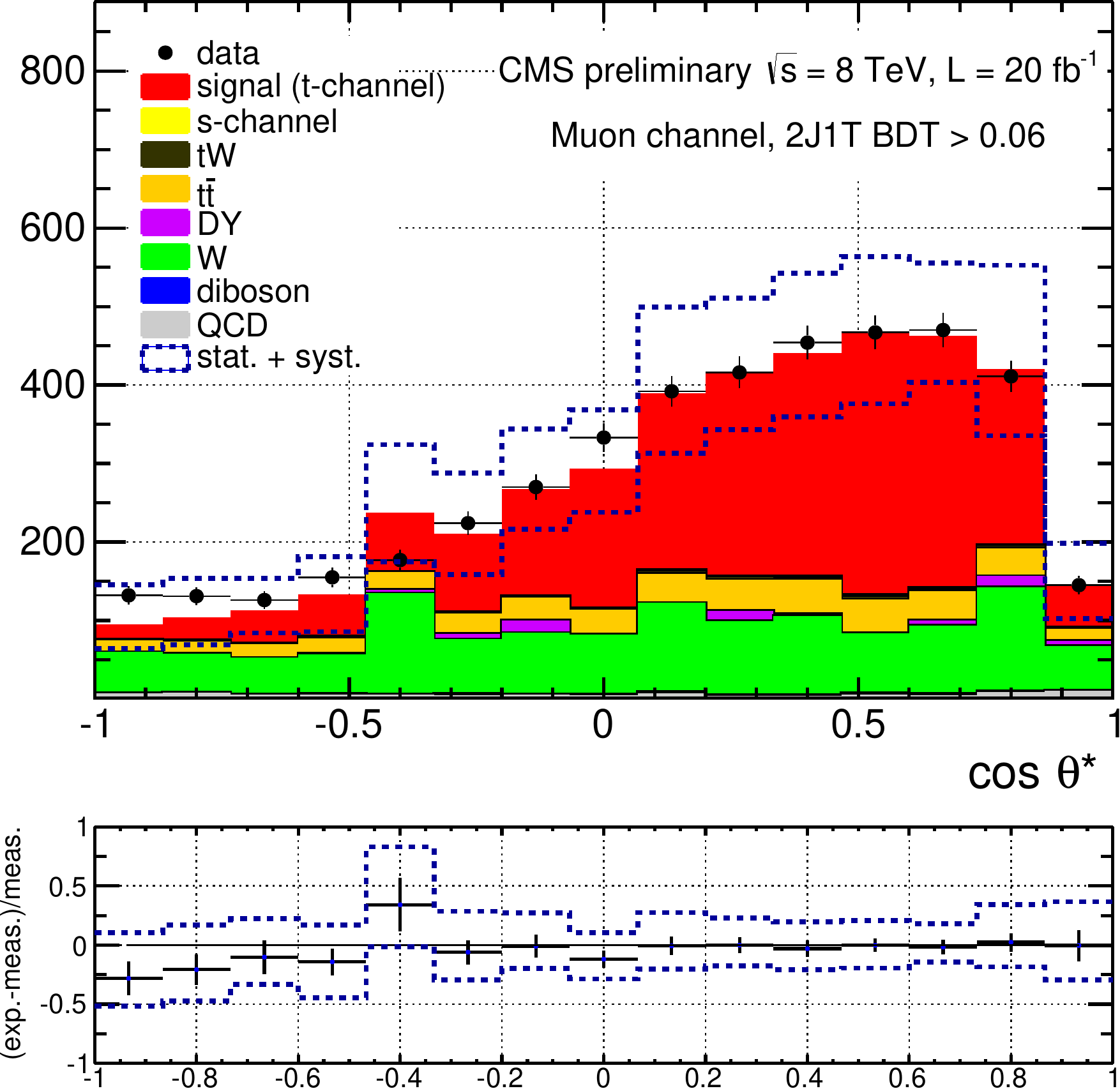}
\center (a) muon channel
\end{minipage}\hspace{1cm}%
\begin{minipage}{7cm}
\includegraphics[height=6.0cm]{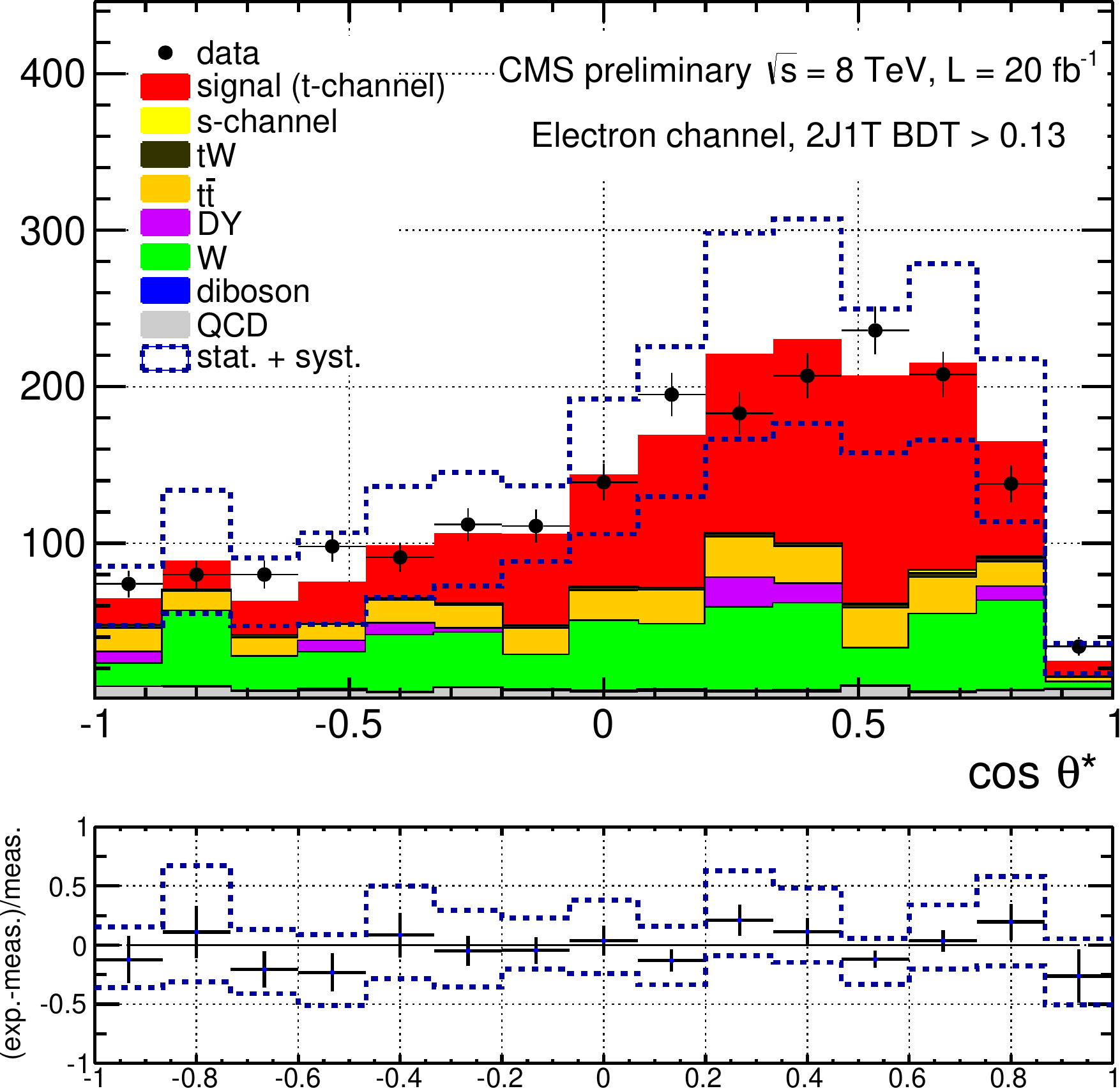}
\center (b) electron channel
\end{minipage} 
\caption{\label{fig:recocostheta}$\costheta$ distributions in signal-enhanced phase space.}
\vspace{-0.7cm}
\end{center}
\end{figure}

In order to extract the $\costheta$ distribution from data, first a top quark candidate is reconstructed by requiring a fixed reconstructed W boson mass and solving the neutrino $p_{z}$ momentum accordingly. Then, the lepton and light jet momentum are boosted into the top quark candidate rest frame. Figure~\ref{fig:recocostheta} shows the resulting $\costheta$ distribution for all events in the signal-enhanced phase space defined by $\BDT>0.06$ and $\BDT>0.13$ for the muon and electron channel respectively.

\section{Unfolding and statistical evaluation}
An unfolding method based on regularized matrix inversion~\cite{tunfold} is deployed to infer the angular distribution and its asymmetry at parton level from the reconstructed $\costheta$ distribution. Prior to unfolding, the data shape is corrected by subtracting the fitted background contributions.

Experimental and theoretical sources of systematics are propagated to the unfolded distribution by repeating the ML fit, background subtraction, and unfolding with modified simulated templates. The analysis accounts for the following sources of systematics: limited number of simulated events, single top t-channel modeling by comparing \textsc{PowHeg} with \textsc{CompHep}, choice of the factorization and renormalization $Q^{2}$-scale, top quark mass, simulation of $\wjets$ events by comparing \textsc{MadGraph} with \textsc{Sherpa}, modeling of the  top quark $\pT$ spectrum in the simulation of $\ttbar$ events, choice of the matching threshold, parton distribution functions, lepton trigger, identification and isolation efficiencies, uncertainty on detector-related jet and $\met$ responses, b-tagging and mis-tagging efficiencies, expected number of pileup interactions, shape and yield of the estimated QCD background from data, ML-fit uncertainties, and the unfolding bias estimated from a linearity check.

\section{Result}

\begin{figure}[h]
\begin{center}
\begin{minipage}{7cm}
\includegraphics[height=5.5cm]{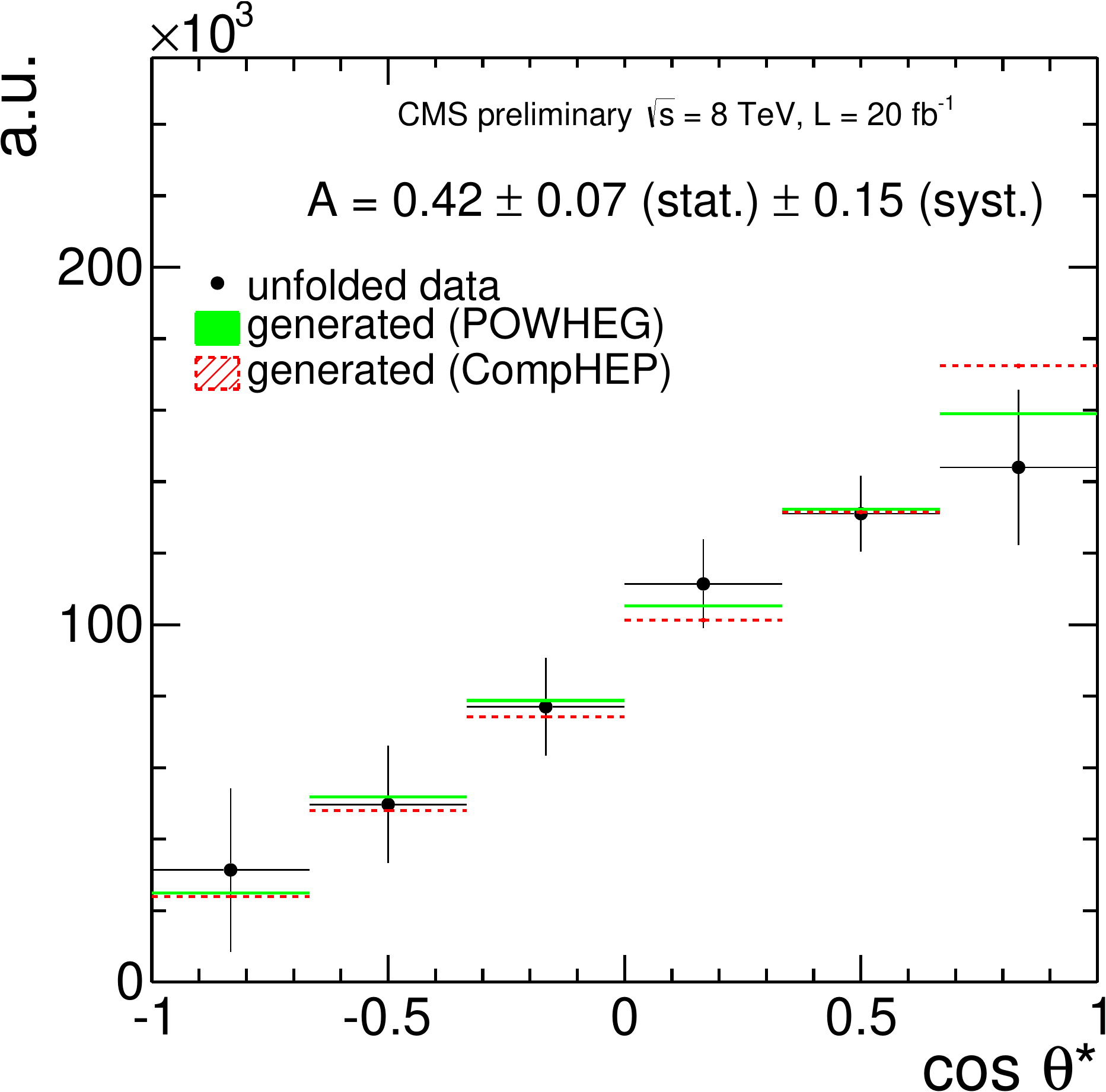}
\center (a) muon channel
\end{minipage}\hspace{1cm}%
\begin{minipage}{7cm}
\includegraphics[height=5.5cm]{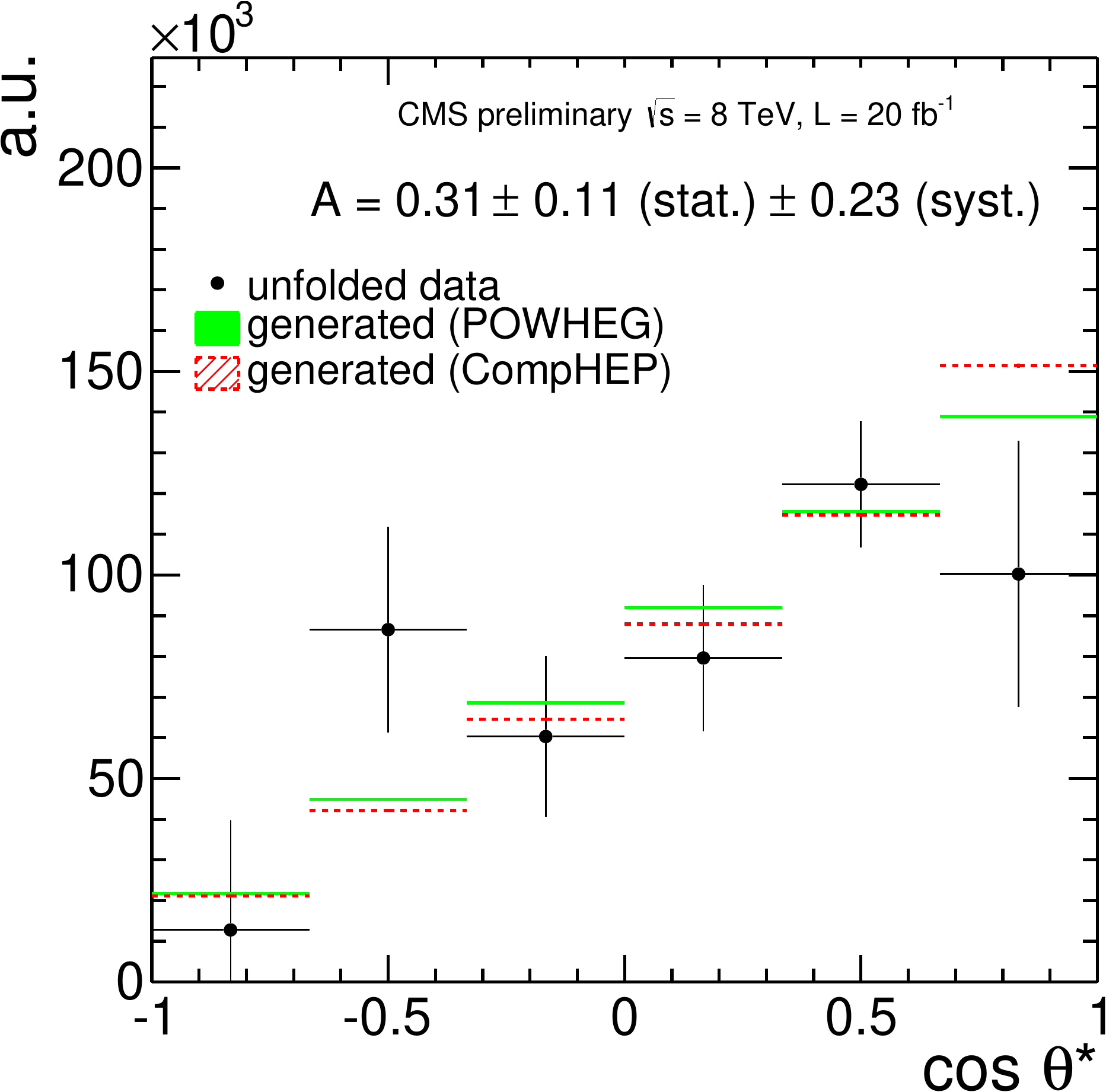}
\center (b) electron channel
\end{minipage} 
\caption{\label{fig:unfoldcostheta}Unfolded $\costheta$ at parton level.}
\end{center}
\vspace{-0.7cm}
\end{figure}

Figure~\ref{fig:unfoldcostheta} shows the $\costheta$ distribution after unfolding compared to the expectation from \textsc{PowHeg} and \textsc{CompHep}. We measure
\begin{align}
A_{\mu}&=0.42\pm 0.07 \textrm{(stat.)} \pm 0.15 \textrm{(syst.)} \\
A_{e}&=0.31\pm 0.11 \textrm{(stat.)} \pm 0.23 \textrm{(syst.)} \\
A_{\mathrm{comb.}}&=0.41\pm 0.06 \textrm{(stat.)} \pm 0.16 \textrm{(syst.)} = 0.41\pm 0.17
\end{align}
for the muon, electron channel and their combination using the Best Linear Unbiased Estimator (BLUE) method~\cite{blue}. The polarization can be inferred through equation~\ref{eq:asym} from the measured asymmetry.

\section{Conclusion}
The first measurement of the top quark polarization has been presented based on approximately $20\unit{fb^{-1}}$ of data recorded during pp collisions at $\sqrt{s}=8\unit{TeV}$.
The polarization is measured to be $P_{t}=0.82 \pm 0.34$ under the assumption of Standard Model couplings in the electroweak top quark decay yielding a lepton spin-analyzing power of $\alpha_{l}=1.0$. The measurement is statistically compatible with the Standard Model which displays a high polarization through the electroweak V-A coupling structure in t-channel single top-quark production.

\section{Acknowledgements}
The author thanks the organizers of the \textsc{Top2014} conference for the possibility of a poster presentation. Furthermore, funding is gratefully accepted from Fonds National de Recherche Scientifique, Belgium.


\section*{References}
\begin{thebibliography}{9}
\bibitem{bernreuther} Bernreuther W 2008 {\it J.Phys.} {\bf G35} 083001 
\bibitem{topmass} ATLAS, CDF, CMS, and D0 Collaborations 2014 arXiv:1403.4427 [hep-ex]
\bibitem{jaaswpol} Aguilar-Saavedra J A and Bernabeu J 2010 {\it Nucl.Phys.} {\bf B840} 349-378 
\bibitem{fabian} Bach F and Ohl T 2014 {\it Phys.Rev.} {\bf D90} 074022 
\bibitem{stpol} CMS Collaboration 2013 {\it CMS Physics Analysis Summary} CMS-PAS-TOP-13-001
\bibitem{cms} CMS Collaboration 2008 {\it JINST} {\bf 3} S08004
\bibitem{tc} CMS Collaboration 2013 {\it CMS Physics Analysis Summary} CMS-PAS-BTV-13-001
\bibitem{tunfold} Schmitt S 2012 {\it ArXiv e-prints} physics.data-an/1205.6201
\bibitem{blue} Lyons L, Gibaut D and Clifford P 1988 {\it Nucl. Instr. and Meth.} {\bf A270} 110-117
\end{thebibliography}
\end{document}